\documentclass[twocolumn,twocolappendix]{aastex631}

\usepackage{graphicx}

\usepackage{subfigure}

\usepackage{amsmath}

\usepackage{amssymb}

\usepackage{multirow}

\usepackage[mathscr]{euscript}

\usepackage[para,online,flushleft]{threeparttable}

\usepackage{multirow}

\usepackage{booktabs}

\usepackage{xr}

\usepackage{array}

\usepackage{graphicx}

\usepackage{tabularx,booktabs}

\usepackage{colortbl}

\usepackage{xcolor}

\newcolumntype{L}[1]{>{\raggedleft\arraybackslash}m{#1}}

\def\sun{$_\odot$}

\begin{document}

\title{Imaging of HH80-81 jet in the NIR shock tracers H$_2$ and [Fe~II]}

\author{Sreelekshmi Mohan}
\affiliation{Indian Institute of Space Science and Technology \\
Thiruvananthapuram 695 547\\
India}

\author{S. Vig}
\affiliation{Indian Institute of Space Science and Technology \\
Thiruvananthapuram 695 547\\
India}

\author{W. P. Varricatt}
\affiliation{UKIRT Observatory \\
University of Hawaii Institute for Astronomy\\
640 N. Aohoku Place, Hilo, HI 96720, USA}

\author{A. Tej}
\affiliation{Indian Institute of Space Science and Technology \\
Thiruvananthapuram 695 547\\
India}





\begin{abstract}
The HH80-81 system is one of the most powerful jets driven by a massive protostar. We present new near-infrared (NIR) line imaging observations of the HH80-81 jet in the H$_2$ (2.122~$\mu$m) and [Fe~II] (1.644~$\mu$m) lines. These lines trace not only the jet close to the exciting source but also the knots located farther away. We have detected nine groups of knot-like structures in the jet including HH80 and HH81 spaced $0.2-0.9$~pc apart. The knots in the northern arm of the jet show only [Fe~II] emission closer to the exciting source, a combination of [Fe~II] and H$_2$ at intermediate distances, and solely H$_2$ emission farther outwards. Towards the southern arm, all the knots exhibit both H$_2$ and [Fe~II] emission. The nature of the shocks is inferred by assimilating the NIR observations with radio and X-ray observations from literature. In the northern arm, we infer the presence of strong dissociative shocks, in the knots located close to the exciting source. The knots in the southern arm that include HH80 and HH81 are explicable as a combination of strong and weak shocks. The mass-loss rates of the knots determined from [Fe~II] luminosities are in the range $\sim 3.0\times 10^{-7}-5.2\times 10^{-5}$~M$_{\odot}$ yr$^{-1}$, consistent with those from massive protostars. Towards the central region, close to the driving source of the jet, we have observed various arcs in H$_2$ emission which resemble bow shocks, and strings of H$_2$ knots which reveal traces of multiple outflows. 
\end{abstract}

\section{INTRODUCTION}

Mass accretion followed by the ejection of material via jets/outflows is an inevitable process during the early stages of star formation. The latter is essential for the removal of excess angular momentum from the system \citep{1986ApJ...301..571P, 1990RvMA....3..234C}. The impact of these supersonic jets on the ambient medium results in the formation of shocks \citep{1997IAUS..182..181H, 2015ApJ...815...96S}. Studies have shown that the shocked plasma is rich in forbidden line emission ([Fe~II], [SiII], [NeII], [OI]) in the optical and infrared wavelengths \citep{Sacco_2012,2009ApJ...692....1D,2005A&A...441..159N,1991ApJ...371..226S}. These lines can assist us in developing a better understanding of the physical parameters of the jet and its dynamics \citep{1999A&A...342..717B,1997A&A...327..671L}. Although high-velocity jets from young stellar objects (YSOs) produce shock-excited optical nebulae at large distances called Herbig-Haro (HH) objects, a search for jet signatures closer to the YSO could help us probe the inner regions. This in turn is essential for the comprehension of jet launching and collimation mechanisms from protostars. Most YSOs are embedded within their molecular clouds causing these innermost regions to be highly obscured from our direct view in optical wavelengths. They can, therefore, be best studied at infrared and radio wavelengths \citep{bally2007observations,2018A&ARv..26....3A}.

Generally, jets are observed to be in the form of a chain of knots rather than continuous emission. Broadly, two hypotheses have been proposed to understand the formation mechanism of knots. In the first scenario, these knots could be formed due to oblique shocks that develop from hydrodynamic, magneto-hydrodynamic (MHD) or thermal instabilities. Simulations and laboratory studies have shown that MHD instabilities could result in jet flow variability since the launch of jets occurs through magneto-centrifugal processes \citep{Huarte_Espinosa_2012, 10.1111/j.1365-2966.2005.09132.x}. In the alternate hypothesis, the variability in the jet flow could also be a result of time variability in the accretion disk \citep{1995ApJ...452..736H}. This time variability could cause episodic ejections of dense jet material which results in the observed knots. Simulations with constant ejection velocity and episodic ejections of material can produce a chain of knot-like material consistent with the observations \citep{2004ApJ...606..483L}.

Observationally, jets have been probed from X-ray to sub-millimeter (sub-mm) \citep{2007A&A...468..515G,2019NatCo..10.3630F,2015Natur.527...70P} and embedded jets can be well probed at NIR wavelengths as they provide dual advantages of lower extinction and good spatial resolution. Numerous studies have been conducted in this domain over the past years \citep{2004A&A...417..247W, 2005A&A...441..159N, 2006ApJ...639..969D, 2006A&A...456..189P,2010MNRAS.404..661V}.
The NIR spectrum is rich in atomic ([Fe~II], [SII], [NI] etc.) as well as molecular (H$_2$) lines \citep{1994ApJ...436..125H, 2000A&A...359.1147E, 2002A&A...393.1035N, 2004A&A...419..999G} with the former tracing the inner ionized  jet and the latter tracing the cold post-shocked molecular gas. The most important NIR lines that have been employed to trace protostellar jets are the [Fe~II] and H$_2$ emission lines \citep{1991ApJ...371..226S,1992A&A...266..439G,1994A&A...289..256S,1995A&A...300..851D,2000AJ....120.1449R}. High resolution observations have shown that the structure of jets/outflows can be divided into two main regions: (i) inner high-velocity material close to the jet axis, and (ii) low velocity less ionized gas near the limb of the jet/outflow \citep{1988ApJ...332..364M, 1990ApJ...362L..25S}. The NIR H$_2$ and [Fe~II] emission lines trace distinct regions along the same jet, and provide information complementary to that of optical and radio observations. Generally, [Fe~II] emission arises from strong shocks in the inner regions close to the jet axis, while H$_2$ is typically associated with the boundary between the jet and the ambient medium \citep{2003A&A...397..693D}. Therefore, a combination of H$_2$ and [Fe~II] imaging of a jet can provide a unique insight into the jet propagation in the molecular cloud. It is noteworthy that there have been large scale surveys targeted on these lines \citep{2014MNRAS.443.2650L,2008MNRAS.387..954D,2009A&A...496..153D,2008MNRAS.387..954D,2009A&A...496..153D,2011MNRAS.413..480F}.

\par In this study, we investigate the HH80-81 jet associated with IRAS~18162-2048 in the NIR (1-5~$\mu$m) wavebands. IRAS~18162-2048 is a massive protostar located at a distance of $1.4$~kpc \citep{2020ApJ...888...41A}. The HH80-81 jet is one of the largest known Herbig-Haro jets, which is also highly collimated extending up to 18.7~pc \citep{2012ApJ...758L..10M}. Proper motion studies of the jet in radio wavelengths have measured velocities as high as $1000$~km~s$^{-1}$ \citep{1995ApJ...449..184M}. The bipolar jet consists of numerous knots and the projected direction of the jet arms lie along the north-east and south-west of the protostar in the sky plane \citep{1989BAAS...21..792R,2001ApJ...562L..91G}. The jet extends up to HH80 in the south-west, and up to HH80N in the north-east. The morphology and kinematics of the HH80-81 jet have been well studied in optical wavelengths by \citet{1998AJ....116.1940H}. These authors estimated the velocities of knots to be up to 600-700~km~s$^{-1}$. Spitzer observations at 8$\mu$m exhibit a bi-conical outflow cavity towards the inner radio jet \citep{2008ApJ...685.1005Q}. Multiple molecular outflows were also detected towards this region indicating active star formation \citep{Fern_ndez_L_pez_2013,1989ApJ...347..894Y,2009ApJ...702L..66Q,2019ApJ...871..141Q}. 
Towards the central region, 25 millimeter (mm) cores have been detected by \citet{2019A&A...623L...8B} using the Atacama Large Millimeter Array (ALMA). Of these, MM1 and MM2 have been found to be the most massive and drivers of outflows in the region, with the HH80-81 jet being excited by MM1 \citep{2019A&A...623L...8B}.

\par Here, we present NIR narrow and broad-band imaging observations of a large extent of the jet ($\sim5$~pc in projection) aimed at characterizing the type of shocks at play. The organization of the paper is as follows. The observation details are presented in Sect.~\ref{data} and the results are described in Sect.~\ref{results}. The results are discussed in Sect.~\ref{interp} and our conclusions are summarised in Sect.~\ref{summary}.

\begin{figure*}
\vspace*{-0.4cm}
\centering
	\includegraphics[scale = 0.75]{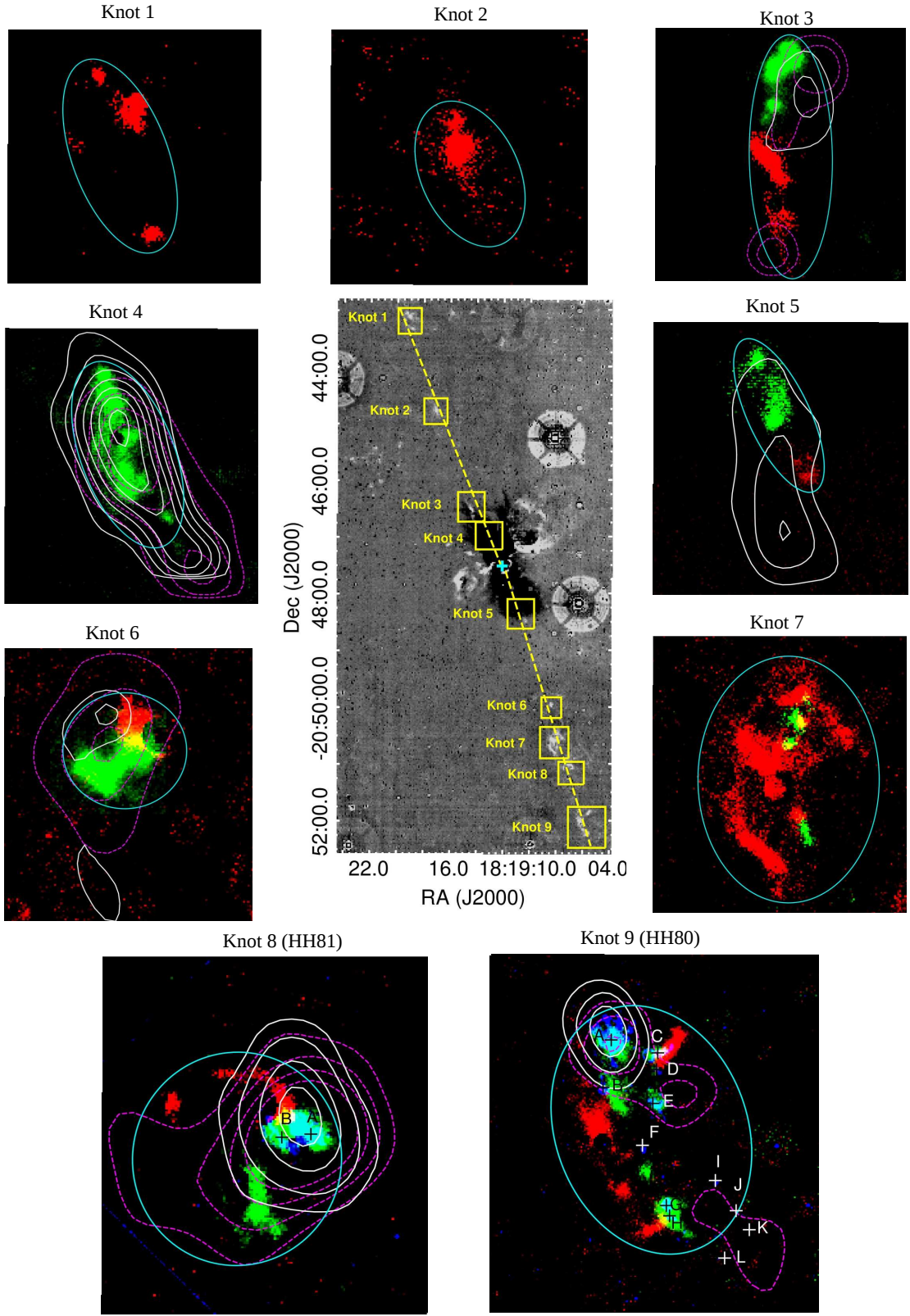} %
    \caption{[Central panel]: Continuum subtracted H$_2$ image of the full length of HH80-81 jet, and [surrounding panels]: color composite image of the individual knots of the jet in the NIR H$_2$ emission (red), [Fe~II] emission (green) and HST optical H$\alpha$+[NII] (blue). The HST observations were carried out only towards the knots HH81 and HH80. Various knots are enclosed in yellow boxes in the central image to show their locations. The details of the locations and sizes of the sub-images are listed in Table.~\ref{tab:table_ref}. The cyan ellipses show the apertures used for the calculation of fluxes of each of the knots. The white contours in the enlarged images of the knots represent VLA 20~cm data obtained from \citet{1993ApJ...416..208M}. The magenta contours (dashed lines) show the GMRT 610~MHz emission from \citep{2018MNRAS.474.3808V}. The cyan cross between knots 4 and 5 in the central figure indicates the driving source of the jet (MM1). Sub-knots towards HH80 and HH81 are shown as crosses and labeled according to \citet{1998AJ....116.1940H}.}
    \label{fig:color_comp_full}
\end{figure*}

\section{OBSERVATIONS AND ARCHIVAL DATA} \label{data}
\subsection{NIR imaging observations}

\par We imaged the IRAS~18162-2048 region using the Wide-Field Camera \citep[WFCAM; ][]{2007A&A...467..777C} mounted on the 3.8~m United Kingdom Infrared Telescope (UKIRT). The WFCAM has four 2048$\times$2048 HgCdTe Hawaii-II arrays with an optical system that provides a pixel scale of 0.4$\arcsec$. This provides a total field of view of 13.5$\arcmin \times$13.5$\arcmin$ per array and 0.21~$\deg^{2}$ for all four arrays together. The target was located in the center of one of the four arrays and the data from only that array was used in the current study. The observations were performed by dithering the target to 9 positions on the array and a 2$\times$2 micro-stepping was done. The resulting mosaics have an image scale of 0.2$\arcsec$/pix. Observations were carried out using the NIR broad-band J, H, K filters and the narrow-band H$_2$ and [Fe~II] filters. The narrow-band H$_2$ filter is centered at $\lambda = 2.122$~$\mu$m and the [Fe~II] line filter at $\lambda = 1.644$~$\mu$m with a full width at half maximum (FWHM) of $0.021$~$\mu$m and $0.028$~$\mu$m, respectively. The observations were conducted between 20200220~UT and 20210920~UT.


\begin{table}
\caption{Details of the coordinates of the center as well as the sizes of sub-images shown in Fig.~\ref{fig:color_comp_full}.}
\hspace*{-2.2cm}
\begin{threeparttable}
\begin{tabular}{l c c c c c c }  \hline \hline	
Source name & RA(J2000) & Dec(J2000) & $\delta$RA & $\delta$Dec\\ 
 & ($\alpha$) & ($\delta$) & ($\arcsec$) & ($\arcsec$) \\ 
\hline
Knot~1 & 18:19:19.10 & -20:43:11.4 & 27.2 & 24.4 \\
Knot~2 & 18:19:17.13 & -20:44:47.2 & 27.2 & 24.4 \\
Knot~3 & 18:19:14.37 & -20:46:30.0 & 32.9 & 31.2 \\
Knot~4 & 18:19:12.97 & -20:47:02.4 & 32.4 & 39.5 \\
Knot~5 & 18:19:10.62 & -20:48:27.4 & 32.7 & 35.4 \\
Knot~6 & 18:19:08.45 & -20:50:03.9 & 24.0 & 27.7 \\
Knot~7 & 18:19:08.13 & -20:50:37.6 & 33.1 & 30.0 \\
Knot~8 (HH81) & 18:19:06.91 & -20:51:06.6 & 28.2 & 32.8 \\
Knot~9 (HH80) & 18:19:05.70 & -20:52:02.4 & 41.7 & 45.4 \\

\hline \
\end{tabular}
\label{tab:table_ref}

\end{threeparttable}
\end{table} 

\par The preliminary reduction including the creation of mosaics was carried out by the Cambridge Astronomical Survey Unit (CASU). For the narrow-band H$_2$ and [Fe~II] filters, further reduction including continuum subtraction was carried out using the Starlink software \citep{2014ASPC..485..391C}. We have obtained imaging observations of H$_2$ in four epochs and those of [Fe~II] in six epochs. The total integration time for J, H, K, H$_2$ and [Fe~II] filters were 720~s, 360~s, 180~s, 1440~s and 1440~s, respectively, for each epoch of observation. For each epoch, the H$_2$ and [Fe~II] images were aligned with K and H-band images, respectively, using {\tt WCSALIGN}. Due to variations in the seeing conditions between the narrow-band and broad-band observations, images with smaller point-spread function (PSF) were smoothed to match with the FWHM of the images with larger PSF using the task {\tt GAUSMOOTH}. For continuum subtraction, sufficient number of isolated bright point sources were selected and their flux counts from both the narrow-band and broad-band sky-subtracted images were calculated. The broad-band to narrow-band flux ratio is then computed. For each imaging observation in an epoch, the broad-band image is scaled using the average value of the ratio of these fluxes and subtracted from the narrow-band image to obtain the continuum-subtracted image for that epoch. The continuum-subtracted images of all the epochs were averaged to obtain the final image. 

The flux calibration of the continuum-subtracted images was carried out using the H and K band flux densities of isolated point sources in the field. For this, we first interpolated the broad-band H and K filter flux densities of a few isolated point sources to the central wavelength of the narrow-band filters. These flux densities were multiplied by the FWHM of the narrow-band filters to obtain the total fluxes. We then derived the photometry of the point sources on the normalized narrow-band images to obtain the counts in ADU. This was used to estimate the flux/ADU and this is used to scale the counts in ADU obtained from the continuum-subtracted narrow-band images to estimate the fluxes. The fluxes are not corrected for extinction as the extinction value towards each knot is not known. 

\par In addition to this, the L$^\prime$ (3.678~$\mu$m) and M$^\prime$ (4.686~$\mu$m) imaging observations were carried out using the UKIRT 1-5~$\mu$m Imager Spectrometer \citep[UIST; ][]{2004SPIE.5492.1160R}. UIST has a 1024 $\times$ 1024 InSb array. The observations were carried out using the 0.12$\arcsec$ pixel$^{-1}$ camera. The object was dithered to four positions along the corners of a square of 40$\arcsec$ side. The total integration time for the L$^\prime$ and M$^\prime$ observations were 160~s and 280~s, respectively. For each setting, a dark observation was done, which was followed by the science observations. The dark subtracted science frames were median combined to create sky flats. Frames in individual pairs were subtracted from each other and divided by the flats. This resulted in a positive and negative beam in the mutually subtracted, flat fielded image. The final mosaic was constructed by combining the positive and negative beams.

\subsection{Optical archival data}

The optical emission from the HH80-81 jet in the filters H$\alpha$ + [NII], [SII] and [OIII] has been imaged using the Wide Field and Planetary Camera 2 (WFPC2) on-board the \textit{Hubble Space Telescope} (HST). These high angular resolution images have been published by \citet{1998AJ....116.1940H}. We have extracted these images from the Hubble Legacy Archive. 
We have employed these images to compare the morphologies of HH80 and HH81 with those from our NIR observations, in order to characterize the nature of shocks.

\begin{figure*}
\hspace*{-1.7cm}
\centering
	\includegraphics[trim={0cm 0cm 0cm 0cm},clip, scale = 0.6]{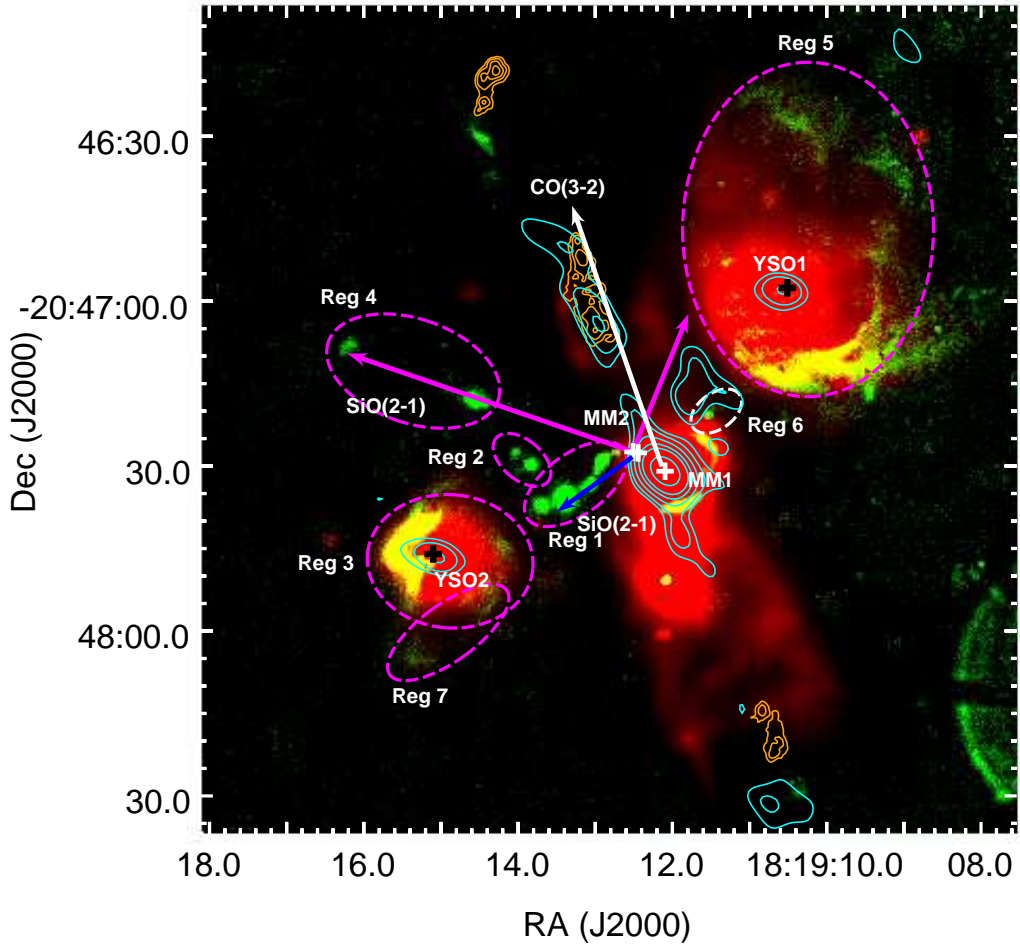} %
    \caption{Two color-composite image of the central region of HH80-81 jet. 
    	The spitzer 8~$\mu$m broad-band image is shown in red and the NIR H$_2$ emission is shown in green. The NIR [Fe~II] emission is overlaid as orange contours, and the arrows depict outflows detected in this region by \citet{2013ApJ...778...72F} and \citet{2019ApJ...871..141Q}. The white crosses indicate two mm cores which are believed to drive some of the detected outflows. The cyan contours show ionized gas at 1300~MHz \citep{2018MNRAS.474.3808V} and the beam
    	is shown towards the bottom left of the image as a white ellipse. The dashed ellipses (magenta) labeled as Reg~$1-7$ are discussed in the text. 
    	The black crosses indicate two YSOs of interest in the region.} 
    \label{H2_spitzer}
\end{figure*}

\begin{figure*}
\hspace*{-2cm}
\centering
	\includegraphics[trim={0cm 0cm 0cm 0cm},clip, scale = 0.6]{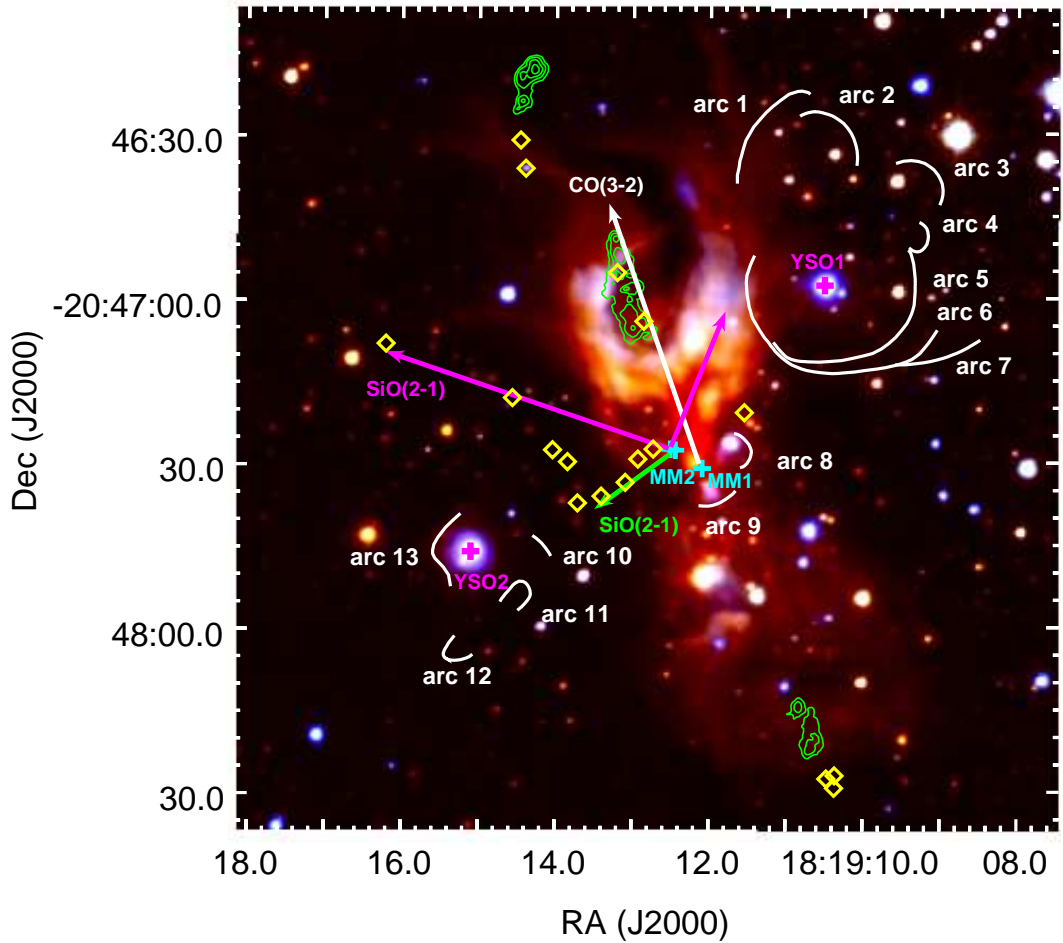} %
    \caption{\textit{JHK} color-composite image of the central region of HH80-81 jet. K is shown in red, H in green and J in blue.
    	The NIR [Fe~II] emission is overlaid as green contours, the arrows depict outflows detected in this region by \citet{2013ApJ...778...72F} and \citet{2019ApJ...871..141Q}. The cyan crosses indicate two mm cores which are believed to drive some of the detected outflows, the magenta crosses indicate two YSOs of interest in the region, the white arcs represent the bow shocks identified in the H$_2$ image of the central region shown in Fig.~\ref{H2_spitzer} and the yellow diamonds show H$_2$ knots.}
    \label{fig:color_comp}
\end{figure*}


\begin{figure*}
\minipage{0.3\textwidth}
\hspace*{-1cm}
  \includegraphics[trim={0cm 0.1cm 0cm 0cm},clip,scale = 0.5]{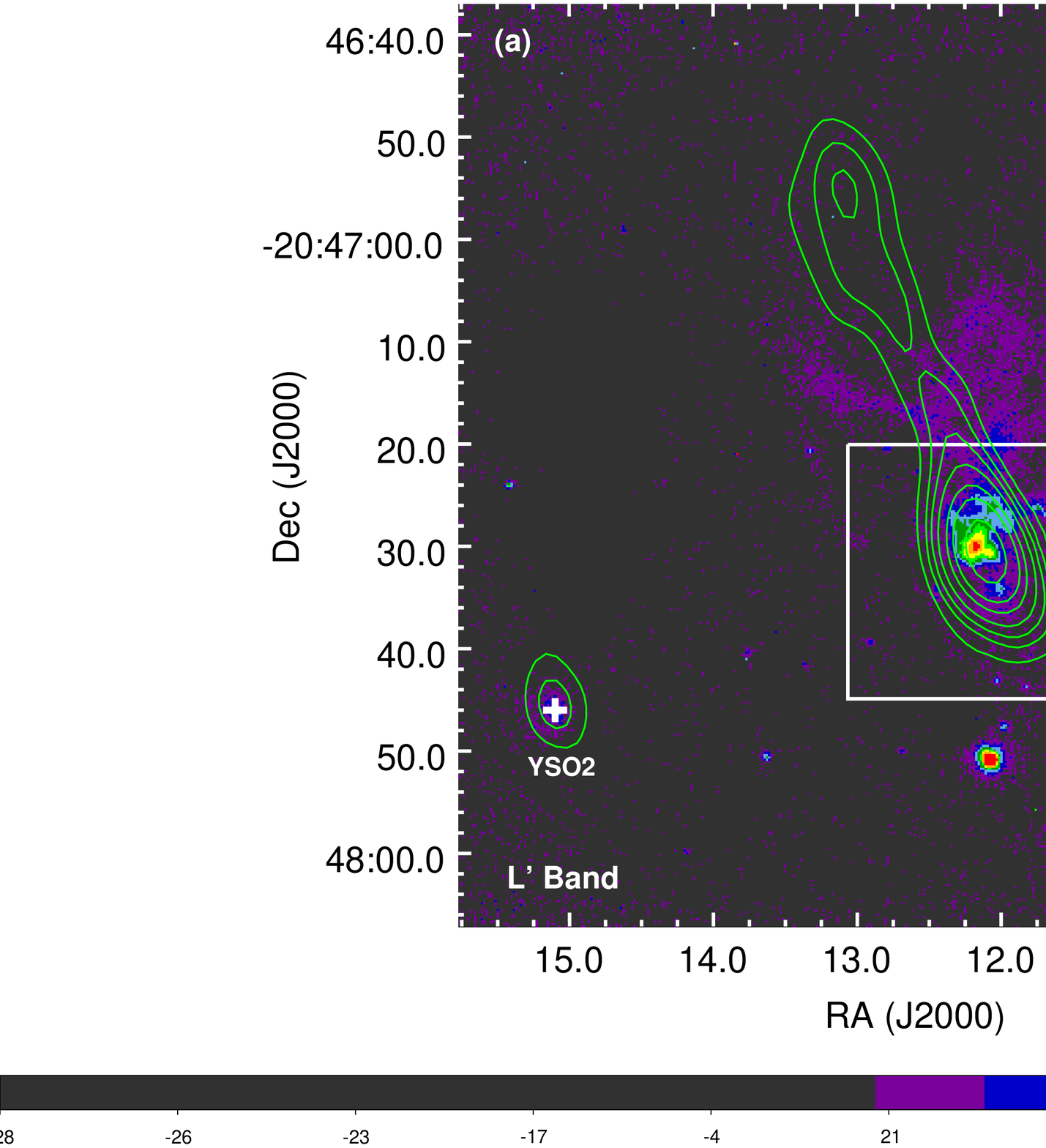}
  \endminipage
\minipage{0.3\textwidth}
\hspace*{6cm}
  \includegraphics[trim={0cm 0.1cm 0cm 0cm},clip,scale = 0.3]{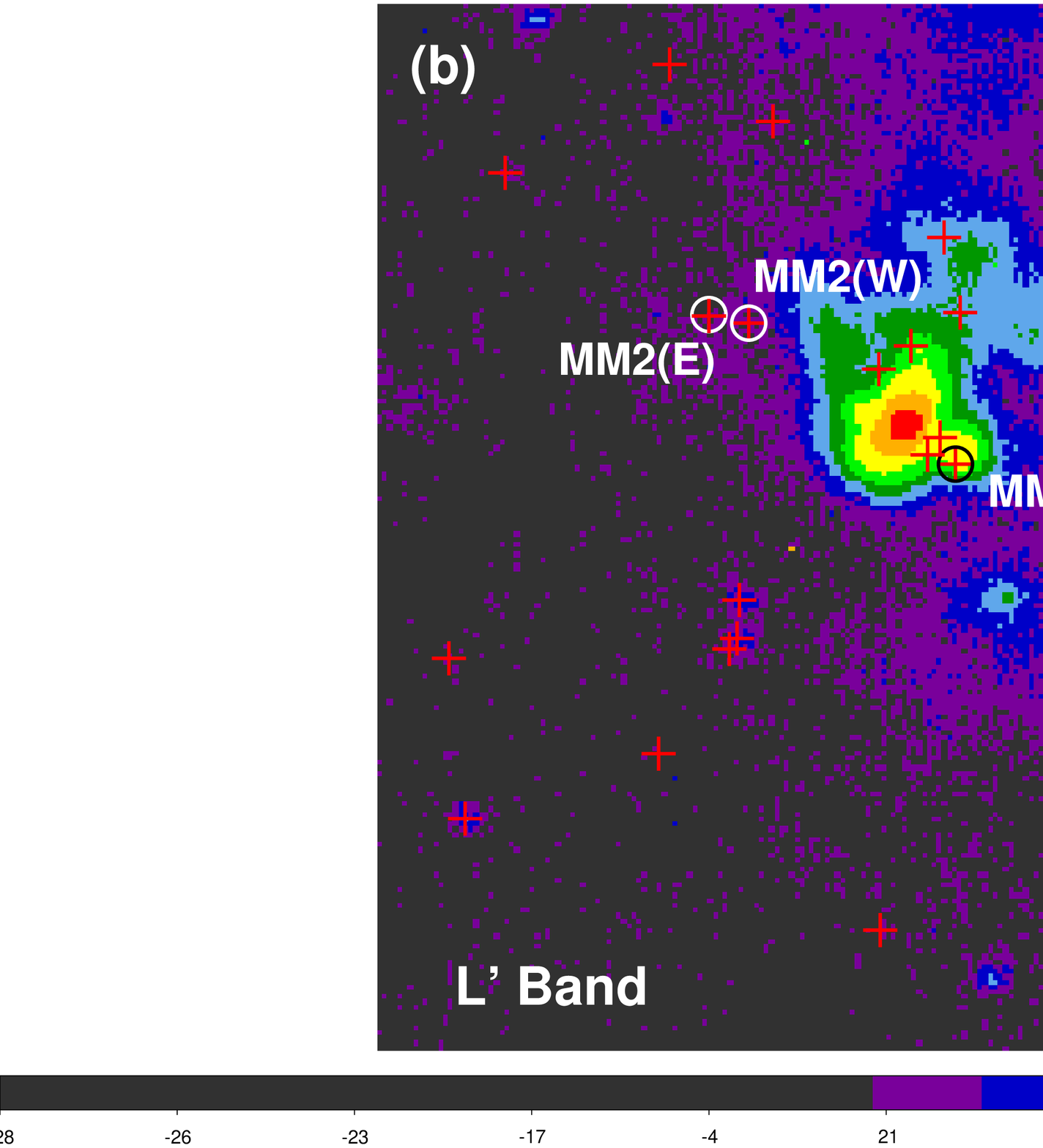}\\
  \hspace*{6cm}
  \includegraphics[trim={0cm 0.1cm 0cm 0cm},clip,scale = 0.345]{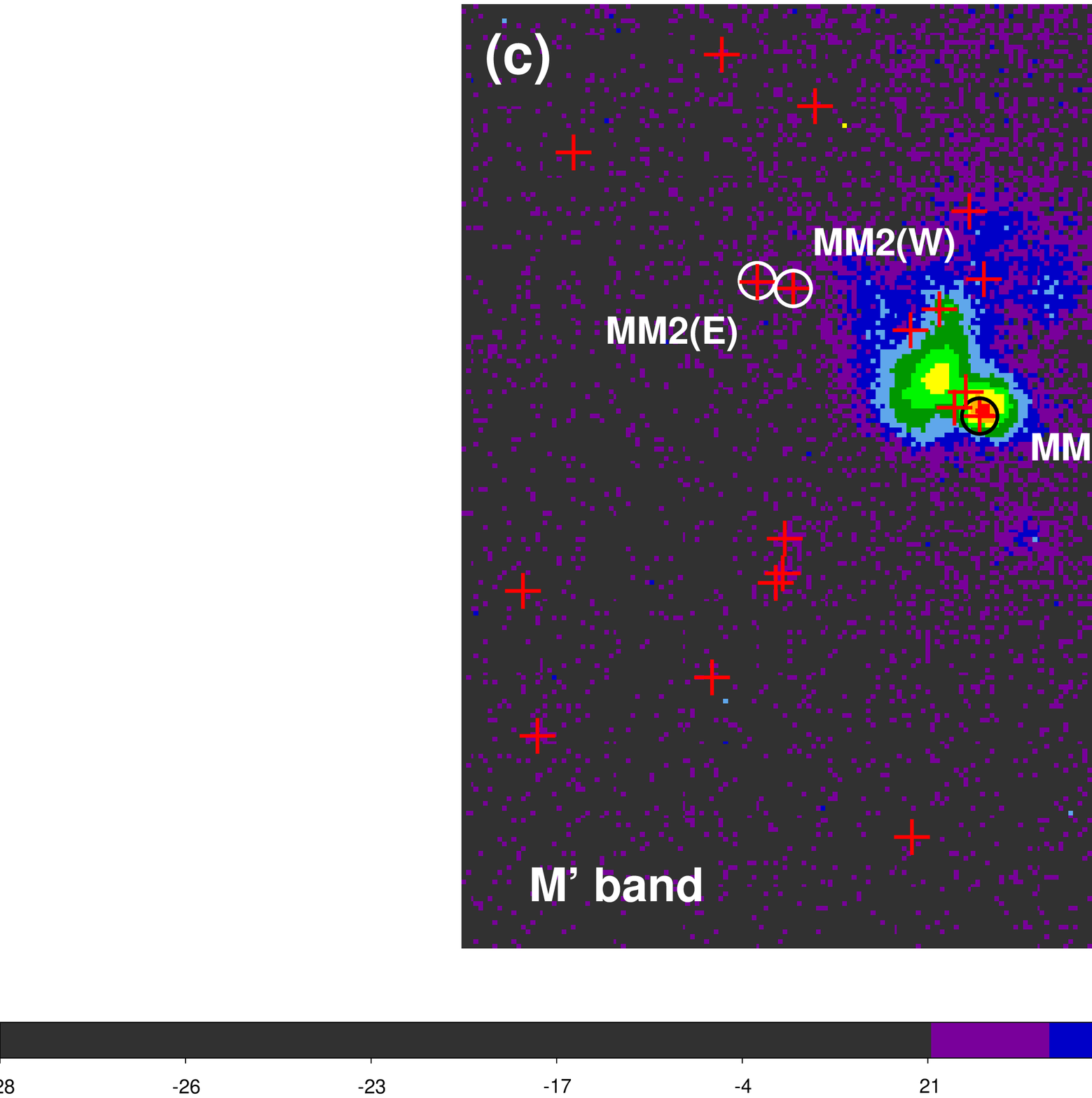}
\endminipage

\caption{(a) L$^\prime$-band image of the central region of HH80-81 jet system
	obtained with UIST. The white box encloses the region of size 0.4~$\arcmin$ $\times$ 0.4~$\arcmin$ shown in (b) L$^\prime$-band emission and, (c) M$^\prime$-band emission. The green contours represent the 20~cm radio continuum emission \citep{1993ApJ...416..208M}, with the beam shown as a white ellipse towards the right corner.	
	The crosses in (b) mark the locations of mm cores in the central region identified by \citet{2019A&A...623L...8B}. The massive cores MM1 and MM2 are labeled.}
\label{fig:LandM_image}
\end{figure*}

\section{RESULTS} \label{results}

\subsection{Line emission in NIR and optical}

The full length of HH80-81 jet as observed in H$_2$ emission is shown in the central panel of Fig.~\ref{fig:color_comp_full}. The surrounding panels display the emission towards the jet in H$_2$, [Fe~II] and the optical H$\alpha$ + [NII] filters in red, green and blue, respectively, revealing the morphologies of individual knots. The details of the sizes and locations of the sub-images are listed in Table.~\ref{tab:table_ref}. We observe that in the $12'$ image obtained by us, the jet displays a chain of knots extending up to 1.9~pc in the north-east and 2~pc in the south-west directions, as projected on the sky plane. Considering the inclination angle of the jet to be $56^\circ$ \citep{1998AJ....116.1940H} from the sky plane, the true extent of the bipolar jet as imaged in NIR is $\sim7$~pc. It is possible that there are additional knots extending farther away from this towards HH80N. We identify a total of nine new knots along the HH80-81 main jet direction and multiple arcs in H$_2$ emission. Four knots are in the northern arm and five are in the southern arm of the jet. We assign numbers 1-9 to the knots as seen in our NIR images from north to south. The Knots~8 and 9 correspond to HH81 and HH80, respectively. We describe a few morphological features of the knots below:
\begin{enumerate}
\item Knots~1 and 2 exhibit only H$_2$ emission, Knot~4 is revealed in only [Fe~II] emission, while the remaining knots exhibit emission in both [Fe~II] and H$_2$ transitions.
\item Knots~1-6 are elongated along the direction of the jet. Knots~7-9 appear broken and have complex morphologies.
\item In none of the knots, the regions corresponding to [Fe~II] and H$_2$ emission completely overlap. In Knots~3 and 6, the [Fe~II] emission clearly lies ahead of the H$_2$ emission along the outward jet direction; whereas, in the case of Knot~5, the reverse is seen.
\item The NIR line emission for Knots~8 and 9 do not coincide with optical line emission. For other knots, the optical data is not available. 
\end{enumerate}

The central region close to the driving source of the jet is shown in Fig.~\ref{H2_spitzer}. The Spitzer 8~$\mu$m broad-band image is shown in red and the H$_2$ emission is depicted in green. The maps are overlaid with the contours of [Fe~II] emission in orange. The association with radio emission is illustrated by contours (cyan) corresponding to the 610~MHz emission \citep{2018MNRAS.474.3808V}. We observe that the emission in H$_2$ is widely distributed across the image, unlike the [Fe~II] emission. It is also evident that the [Fe~II] emission arises from the highly ionized inner jet close to the axis while the mid-infrared (MIR) emission delineates the outflow cavity. Multiple bow shocks and knots are observed in H$_2$ and the majority of this emission appears to be distributed in directions away from the HH80-81 jet axis. The radio jet is well delineated by the [Fe~II] emission. 

The prominent H$_2$ features in the region are:
\begin{enumerate}
\item The HH80-81 main jet: This extends along the north-east $-$ south-west direction. The emission is seen as knots extending along the jet as well as arcs close to the central region. A detailed discussion of this jet is presented in Sect.~\ref{interp}. The catalogue number of the knots associated with this jet in the Catalogue of Molecular Hydrogen Emission-Line Objects (MHOs) in Outflows from Young Stars \citep{2010A&A...511A..24D} is MHO~2354.
\item Reg~1: This comprises a chain of 5 H$_2$ knots towards the eastern side, that are enclosed within a dashed ellipse shown in Fig.~\ref{H2_spitzer}. 
\item Reg~2: This includes two distinct H$_2$ knots to the north-east of Reg~1.
\item Reg~3: This region includes arcs of H$_2$ emission along the east and west directions on either side of a YSO \citep{2008ApJ...685.1005Q}, as shown in Fig.~\ref{H2_spitzer}, facing inwards of the dashed ellipse of size $0.5'\times0.4'$. The arc towards the east is very strong in emission and with overlapping emission at 8~$\mu$m, it is seen as yellow in the figure. We note that the nebulosity at 8~$\mu$m appears well enclosed within the defined ellipse. 
\item Reg~4: This region shows two knots aligned north-east with respect to the central region.
\item Reg~5: The H$_2$ emission shows arc-shaped patterns over a region of size $1.0'\times0.7'$. The arcs appear to surround the nebulosity observed at 8~$\mu$m. The emission to the south of the enclosed ellipse shows strong H$_2$ emission, overlapping with 8~$\mu$m emission. 
\item Reg~6: This region includes a knot towards the west of the central region. 
\item Reg~7: Two faint arcs in H$_2$ are observed facing each other within the region enclosed by the ellipse in this region. 
\end{enumerate}

\begin{table*}
\setlength{\tabcolsep}{16pt}
\caption{Fluxes of HH80-81 jet knots in the narrow-band H$_2$ and [Fe~II] filters, the sizes of apertures (semi-major $\times$ semi-minor) used for the flux calculation, and the [Fe~II]/H$_2$ line ratios ($\mathscr{R}$) for the HH80-81 main jet knots in which both H$_2$ and [Fe~II] emission were detected.}
\begin{center}
\hspace*{-2.2cm}
\begin{tabular}{l c c c c c c }  \hline \hline	
Source name & H$_2$ flux &  [Fe~II] flux & Aperture size & $\mathscr{R}$ = [Fe~II]/H$_2$ \\
& (  10$^{-14}$  erg s$^{-1}$ cm$^{-2}$) &  ( 10$^{-14}$   erg s$^{-1}$ cm$^{-2}$) &  & \\ 
\hline
Knot~1 &   1.7 & $-$ $^\dagger$ &  10.7$\arcsec$ $\times$ 4.7$\arcsec$ &  $-$ \\
Knot~2 &   1.3 & $-$ $^\dagger$ & 8.1$\arcsec$ $\times$ 4.7$\arcsec$ & $-$  \\
Knot~3 &   2.6 &   4.6 & 14.8$\arcsec$ $\times$ 5.5$\arcsec$ &  1.8\\
Knot~4 & $-$ $^{\ddagger}$ &   11.7 & 13.9$\arcsec$ $\times$ 6.0$\arcsec$ & $-$  \\
Knot~5 &   0.4 &   4.6 & 10.7$\arcsec$ $\times$ 3.9$\arcsec$ &  12.1 \\
Knot~6 &   0.9 &    5.2 & 5.9$\arcsec$ $\times$ 5.8$\arcsec$ &  6.0  \\
Knot~7 &   5.2 &    3.1 & 14.6$\arcsec$ $\times$ 10.7$\arcsec$ & 0.6  \\
Knot~8 &   1.5 &    7.0 & 9.6$\arcsec$ $\times$ 9.1$\arcsec$ &  4.7  \\
Knot~9 &   4.3 &    13.6 & 16.2$\arcsec$ $\times$ 12.1$\arcsec$ &  3.2  \\

\hline \
\end{tabular}
\label{tab:flux}
\tablecomments{$^\dagger$ [Fe~II] emission is not detected in these knots, $^{\ddagger}$ H$_2$ emission is not detected in this knot.}
\end{center}
\end{table*}

The MHO catalogue numbers of Regs~1-7 are MHO~2355$-$2361. A detailed description of the morphology of knots in narrow line emission bands is presented in Appendix~A. 
	
We have estimated the integrated fluxes of knots in the narrow-band H$_2$ and [Fe~II] filters by using elliptical apertures for each of the knots, which are marked in cyan in Fig.~\ref{fig:color_comp_full}. The knot fluxes along with the sizes of apertures (semi-major $\times$ semi-minor) used for the flux calculation are tabulated in Table.~\ref{tab:flux}. The H$_2$ and [Fe~II] fluxes of the knots are in the range $0.4 - 5.2\times$10$^{-14}$ erg s$^{-1}$ cm$^{-2}$ and $3.1 - 13.6\times$10$^{-14}$ erg s$^{-1}$ cm$^{-2}$, respectively. Knots~1 and 2 do not have any associated [Fe~II] emission. We have also calculated the ratio of [Fe~II]/H$_2$ emission towards knots where both are detected. This ratio is represented as $\mathscr{R}$, and estimated in order to understand the nature of shock across different knots. These values are also tabulated in Table~\ref{tab:flux} and are comparable with those obtained from observational studies conducted towards other YSO jet sources \citep{2002ApJ...564..839L, 2010A&A...511A...5G, 2004MNRAS.353..813M}. It is worthwhile to note that the extinction can be quite non-uniform along the length of the jet, and will also be affected by the inclination of the jet with respect to the line-of-sight (LOS). However, having stated this, we observe that $\mathscr{R}$ varies widely across the knots, from 0.6 in Knot~7 suggesting relatively low excitation of [Fe~II] emission to 12.1 in Knot~5 where [Fe~II] emission clearly dominates over the H$_2$ emission. This clearly indicates that there is a considerable difference in the excitation conditions of the knots. We discuss this in detail in Sect.~\ref{interp}.

\subsection{Broad-band emission in NIR and MIR}

In this section, we describe the emission from the central region in NIR and MIR broad-band filters. The emission through the NIR J, H and K bands can be seen in Fig.~\ref{fig:color_comp}. The J, H and K band emission towards the northern extension of the jet is up to $\sim1'=0.4$~pc. On the other hand, towards the southern arm, the NIR J and H band emission is observed up to 0.2~pc whereas the K band emission is seen to extend up to 0.3~pc.
We identify a sharp bend in the jet cavity towards the northern arm, where it is initially directed westwards but veers towards the east, see Fig.~\ref{fig:color_comp}. The most massive mm cores observed in this region are MM1 and MM2 \citep{2019A&A...623L...8B}, which are marked in the figure. In addition, the knots and arcs which are discussed in Sect.~3.1 are also shown in the figure.

The emission towards the central region in the L$^\prime$ and M$^\prime$ bands can be viewed in Fig.~\ref{fig:LandM_image}. We note that the cavity is not seen very clearly in these images and we attribute this to the low signal-to-noise ratio images. While the L$^\prime$ band image shows hints of emission towards the jet cavity near the center (see Fig.~\ref{fig:LandM_image}(a)), this is not detected in the M$^\prime$ image. A comparison with MIR IRAC images from the space-based \textit{Spitzer} mission shows the presence of strong MIR emission from the cavity walls (Fig.~\ref{H2_spitzer}). Cavities around jets in MIR have been observed earlier and the emission has been attributed to thermal dust which is directly heated by the central source that drives the outflow \citep{2006ApJ...642L..57D}. We observe a few point-like sources in our moderately high resolution L$^\prime$ and M$^\prime$ images, and we identify a few of them as cores that have been observed using mm ALMA emission \citep{2019A&A...623L...8B}. The brightest emission in L$^\prime$ image is observed at $\sim$2000~au to the north-east of MM1. While this emission is very likely from the outflow cavity, there exists a possibility that this is a YSO considering the point-like morphology seen in the M$^\prime$ image. The lack of associated mm emission gives more credence to the outflow cavity hypothesis although the possibility of an evolved YSO cannot be completely ruled out. MM1 is brighter in the M$^\prime$ band compared to L$^\prime$ band, indicating that it is indeed young.  We have not detected any emission associated with MM2(E) and MM2(W) in the M$^\prime$ band.

\section{DISCUSSION} \label{interp}

We have identified nine knots along the HH80-81 jet: four on the northern side and five on the southern side of the driving source, as shown in Fig.~\ref{fig:color_comp_full}. The morphology of the individual knots are outlined in Appendix~A. In general, the excitation of 2.122~$\mu$m $\nu= 1-0$~S(1) ro-vibrational line of H$_2$ molecule can occur due to either of the following mechanisms: (i) collisional excitation in dense molecular gas which is heated up to a temperature of 2000~K by shock waves (shock-excited), or (ii) absorption of ultraviolet radiation emitted by the young star (fluorescence). In the case of HH80-81 jet, the H$_2$ emission is observed at large distances from the central stellar source (up to 2~pc in projection on each side). Therefore, we suggest that the shock origin is a more plausible scenario than the fluorescence phenomenon. The NIR emission lines of the Fe atom are also tracers of jets \citep{2000AJ....120.1449R, 2003Ap&SS.287...21P, 2013ApJ...778...71G, 10.1093/mnras/stu788}. There are 16 low-lying energy levels in Fe$^+$ ion that can be easily excited. These excitations can therefore be readily achieved in shocks. 

\subsection{Nature of shock in knots} \label{nature}

In this section, we qualitatively characterize the nature of shocks at play in each of the knots. Depending on the strength, structure and nature of the shock generated by jets, the shocked plasma can cool via emission of atomic, molecular or ionic lines. Thus the emission line features can be utilized to uncover the type and nature of shocks that give rise to these emissions. Based on their physical properties, shocks can be broadly classified as (i) Jump shocks (J-type) (ii) Continuous shocks (C-type) and (iii) J-type with magnetic precursor \citep{1980ApJ...241.1021D,2003MNRAS.341...70F}. C-shocks are typically associated with velocities $\sim$40 - 50~km~s$^{-1}$, low ionization of molecular gas and strong magnetic fields. They are rich in infrared emission lines (H$_2$, CO, H$_2$O etc.). J-shocks on the other hand are faster and mostly rich in optical and ultraviolet emission lines and require high ionization levels with low magnetic field strength \citep{1989ApJ...342..306H}. Younger shocks typically tend to be of J-type (strong) and such shocks may eventually evolve into C-type (weak) in the presence of a strong magnetic field. A shock that has not attained thermal equilibrium could display the properties of both C and J-type shocks. The intermediate stage in this evolution is called the J-type shock with a magnetic precursor. While spectroscopy is ideal for probing the intensities of excitation lines and hence the nature of shock at a specific location, the spatial distribution of the different excitation lines through a morphological study is equally important as it provides strong clues about jet propagation in the ambient medium. 

We utilize the H$_2$ and [Fe~II] line images to comment on the possible nature of shocks in different knots of the HH80-81 jet. Molecular H$_2$ is a good tracer of low-velocity ($10-50$~km~s$^{-1}$) weak molecular shocks associated with protostellar jets, i.e. shocks at low velocities \citep{1980ApJ...241.1021D}. In such cases, the dissociation of H$_2$ molecule does not occur. However, in regions of low densities, the H$_2$ molecule remains undissociated even up to $80$~km~s$^{-1}$ \citep{2002MNRAS.332..985L}. Infrared H$_2$ emission could occur either in non-dissociative C-shocks or low velocity J-shocks. This is corroborated by the fact that numerous HH flows \citep{1988ApJ...334L..99S,1989ESOC...33..331L} and several CO outflows have exhibited H$_2$ emission features tracing weaker regions of shocks with low excitation levels. J-shocks with speeds greater than 25~km~s$^{-1}$ are found to dissociate H$_2$ \citep{1977ApJ...216..713K}, whereas C-shocks are capable of accelerating H$_2$ to velocities up to 50-80~km~s$^{-1}$ without dissociating it \citep{1980ApJ...241.1021D,2002MNRAS.332..985L}. [Fe~II] emission is widely used to understand fast and dissociative J-shocks caused by jets with velocities larger than 50~km~s$^{-1}$. At these velocities, the head of the bow shock can destroy the grain in which iron is locked up \citep{2002ApJ...564..839L}. The grain destruction releases iron in the gas phase and this is followed by ionization of the neutral iron through charge transfer reactions with ions. It is possible that J-shocks can have cooler wings, where H$_2$ molecules are excited behind the [Fe~II] emitting region. Alternately, J-shocks with magnetic precursors can give rise to a geometry in which H$_2$ emission is seen in front of the [Fe~II] emission \citep{1980ApJ...241.1021D}. The morphology of emission in shock tracers can provide a resourceful gauge to examine the interaction of the jet with the ambient medium. 

In the HH80-81 jet, from the observed morphology of knots aligned along the jet, we attempt to segregate strong and weak shocks and correlate this information with signatures of C-shocks, J-shocks as well J-shocks with magnetic precursors. 
In Knot~1, the elongated cavity like H$_2$ emission and the absence of [Fe~II] and radio emission towards this distant knot could be explicable by the weakening of the shock strength with radial distance as it moves outwards from the central source. The absence of [Fe~II] and radio emission helps to exclude J-shock as the exciting mechanism at this knot location, and hence we propose that the shock in this knot could be a C-type shock. For Knot~2, although its morphology is different from that of Knot~1, strong H$_2$ emission and the absence of [Fe~II] and radio emission again points towards the possibility of a C-type shock due to weakening of the shock with radial distance from the central source. The filled morphology could suggest the presence of a stronger shock. Compared to Knot~1, this shock is possibly stronger due to smaller radial distance from the driving source. Alternatively, the emission could be from the front side of the wall-cavity as mentioned earlier. 

In Knot~3 the [Fe~II] emission appears to arise from the head of the bow shock and the H$_2$ emission corresponds to the cooler wing of the bow shock i.e interior to that of [Fe~II] emission in the post-shock cooling zone. The fast dissociative shock head of the J-shock results in the generation of Fe$^+$ ions in the immediate post-shock region which undergoes collisional excitation and emission of forbidden [Fe~II] transition. Behind the shock front, molecular re-formation occurs due to which H$_2$ emission is observed \citep{1989ApJ...342..306H}. For the H$_2$ molecule to remain undissociated, the angle between the direction of the oblique planar shock responsible for the emission and the jet direction is expected to be less than 5-7$^\circ$ for a J-type shock and 10-15$^\circ$ for a C-type shock \citep{2000MNRAS.318..747D}. In Knot~3, we find that the angle between the H$_2$ knot and the flow direction is $\sim16^\circ$ and between the [Fe~II] knot and jet is $\sim42^\circ$. This strongly suggests the C-type nature of the cooler wings and J-type nature of the shock head where H$_2$ emission is missing. A similar case is observed towards Knot~6. The [Fe~II] emission is located at the bow head and H$_2$ emission is at the cooler wings behind the shock front. This again points to the possibility of a strong dissociative J-shock with cooler wings. In Knot~4, only [Fe~II] emission is present due to the highly dissociative nature of the shock.

We observe the presence of H$_2$ emission exterior to the [Fe~II] emission in Knot~5, and it is possible that a J-shock with a magnetic precursor can account for this scenario. The magnetic precursor causes heating and compression of the materials upstream of the shock front when the shock wave travels in a weakly ionized gas in the presence of a transverse magnetic field \citep{1989ApJ...347L..31H, 2003MNRAS.341...70F}. However, here the neutrals undergo a discontinuous change of physical conditions across the shock. A low ionization fraction reduces the rate of ion-neutral collisions resulting in weaker compression of the magnetic field. Under these low-ionization conditions, if the magnetic field is strong enough the shock will possess a magnetic precursor which compresses the magnetic field even before the shock front crosses a region. This happens because in this scenario because the magnetosonic waves can propagate faster than the actual shock front \citep{1980ApJ...241.1021D}. Hence, the arrival of shock compression before the shock front at each location along the path of the jet flow provides favorable conditions for excitation of H$_2$ molecules without dissociating it. This could explain the observed H$_2$ emission in this knot. Furthermore, behind the shock compression, at the dissociating shock front, we observe [Fe~II] emission in accordance with predictions from the theory. 

The Herbig-Haro objects HH80 and HH81 display complex morphologies. For HH81 (Knot~8), the H$_2$ and [Fe~II]/optical emission appears to arise from bow shocks which are arched in a direction that is opposite to the jet propagation. The arching of the bow shock opposite to the direction of the flow can be explained by a scenario where the jet encounters a small and dense clump of material on its path \citep{1991ApJ...371..226S}. The bow shock then curves in the direction opposite to the direction of the jet flow. Towards HH80 (Knot~9), multiple knot-like structures are observed. The presence of elongated H$_2$ emission towards the eastern lateral edge of the [Fe~II] emitting region could have an origin in the wiggling or sideways motion of the jet flow \citep{1993ApJ...416..208M, 1994ApJ...422L..91E}. An alternative explanation is provided by \citet{1996ApJ...462..804N} which introduces the grazing jet/cloud core collision scenario. This states that when the jet impacts on the molecular core at an angle, it gets deflected and moves away from the surface of impact. As the jet deflects away, the external pressure forces it back causing the outflow to bounce back and forth while some gas slides closer to the core. This is the zone of interaction between the atomic and molecular gas, and the H$_2$ excitation is most likely to be arising from this region. We suggest that the back-and-forth motion of the jet could create internal shocks that are substantiated by the presence of elongated [Fe~II] emission, located closer to the jet-axis than the H$_2$ emission. The complex morphology of the entire knot could arise from a scenario where thermal instabilities develop in an optically thin jet which ultimately fragments and breaks up the shocked material \citep{1965ApJ...142..531F, 1970ApJ...161..451H}. Proper motion studies aimed at understanding the kinematics of these HH80-81 knots have measured velocities for all the sub-knots, and these lie in the range $74-916$~km/s  \citep{1998AJ....116.1940H}. The direction of velocity is different for different sub-knots, which supports our claim regarding fragmentation of these knots as a result of thermal instabilities. For HH80 and HH81, it is difficult to conclude the nature of shock that explains the entire complex morphology of knots using the NIR and optical alone, and it is possible that one or multiple types of shocks could be generated in the sub-knots.

The $\mathscr{R}$ value of each knot is representative of the strength of the shock generated. The \citet{1989ApJ...342..306H} model predicts that a [Fe~II]/H$_2$ ratio greater than 0.5 is indicative of high-velocity dissociative shocks. The $\mathscr{R}$ values of the HH80-81 jet knots are greater than 0.5 (Table.~\ref{tab:flux}) thereby indicating the presence of J-type shocks throughout the length of the jet. In the case of Knot~5 which is located in the southern arm of the jet, the $\mathscr{R}$ value is significantly high, due to its close proximity to the driving source. Here, the shocks are very strong and highly dissociative, thereby, favoring [Fe~II] emission over H$_2$. Similarly, the corresponding closest knot in the northern arm, Knot~4, is much stronger and dissociative resulting in only [Fe~II] emission.

We, next, utilize the location of soft X-ray emission to discuss the plausible nature of shocks in these regions. The highly dissociative head of a J-shock being the strongest section of the shock is believed to be rich in UV and X-ray emissions. X-ray emission has been detected towards HH80 (sub-knots~A, C and G) and HH81 (sub-knots A) by \citet{rodriguez2019particle} that has overlap with the optical emission. In addition, a scrutiny of the [Fe~II]/optical and soft X-ray images \citep[Fig.~2; ][]{rodriguez2019particle} suggest the X-ray emission to be marginally ahead of the [Fe~II]/optical emission in the outward propagating jet direction. This points towards the likely presence of a J-type shock at these locations. For the other sub-knots, the shock mechanisms are difficult to disentangle.

Although we have tried to understand the nature of shocks, i.e. strong (J), weak (C), or intermediate (J with magnetic precursor), based on the spatial distribution and overlap of H$_2$ and [Fe~II] lines, it is possible that extinction effects could also play a role in regions where H$_2$ is detected but [Fe~II] is not detected as the former suffers lower ($\sim50\%$) extinction compared to the latter. This is discussed in detail in Sect. 4.2.

\begin{table*}
\setlength{\tabcolsep}{8pt}
\caption{Physical parameters of HH80-81 jet knots and, mass-loss and momentum rates for the knots in which [Fe~II] emission was detected.}
\begin{center}
\hspace*{-2.2cm}
\begin{tabular}{l c c c c c c }  \hline \hline	
Source name & L$_{\text{H}_2}$ & L$_{\text{[Fe~II]}}$ & $v_{\perp}$ &  $l_{\perp}$ &$\dot{M}_{\text{jet}}$ &  $\dot{P}$\\
  & (10$^{31}$ erg s$^{-1}$) & (10$^{32}$ erg s$^{-1}$) & km~s$^{-1}$ & ($\arcsec$)  &(10$^{-5}$ M$_{\odot}$ yr$^{-1}$) &  (10$^{-2}$ M$_{\odot}$ yr$^{-1}$ km~s$^{-1}$) \\ 
\hline
Knot~1   & 8.9  & $-^{\dagger}$  & $-^{\dagger}$ & $-^{\dagger}$   &$-^{\dagger}$ &  $-^{\dagger}$\\
Knot~2   & 6.8  & $-^{\dagger}$  & $-^{\dagger}$ & $-^{\dagger}$    &  $-^{\dagger}$ &  $-^{\dagger}$ \\
Knot~3   & 13.2  & 13.7  & 1000 & 11.5  &4.3 &  4.3 \\
Knot~4   & $-^{\ddagger}$ & 34.5  & 1000 &  24.0   &5.2&  5.2\\
Knot~5   & 2.0  & 13.6  &  1000&  14.0    &3.5&  3.5\\
Knot~6   & 1.0  & 1.4  &  500&  6.5   &0.4& 0.2 \\
Knot~7   & 5.7  & 0.8  & 500 &  17.5    &0.1&  0.04\\
Knot~8   & 0.5  & 0.3  &  500 & 11.5   &0.04 &  0.02\\
Knot~9   & 1.3  & 0.5 &  500 &  30.0   &0.03 &  0.02\\

\hline \
\end{tabular}
\label{tab:lum_mass}
\tablecomments{$^\dagger$ [Fe~II] emission is not detected in these knots, $^{\ddagger}$ H$_2$ emission is not detected in this knot}
\end{center}
\end{table*} 

\subsection{HH80-81 jet propagation} \label{propagation}

A comprehensive view of the jet reveals that towards the northern arm, the knots closer to the driving source display solely [Fe~II] and radio emission or a combination of [Fe~II], radio and molecular H$_2$ emission, whereas H$_2$ emission dominates in the farther knots. The H$_2$ emission is believed to trace the diffuse regions in outflow lobes which are not well-collimated whereas [Fe~II] and radio emission reveals the compact jet where the fast moving jet ejecta produces stronger shocks \citep{2002ApJ...564..839L}. This implies that [Fe~II] and radio emission traces the ejecta accelerated directly by the exciting source, whereas H$_2$ emission arises from shocked ambient gas where the ejecta interacts with it. The [Fe~II] emission gradually diminishes with radial distance from the exciting source towards this arm. From this, we can infer that at smaller radial distances from the driving source, shocks are stronger J-type, and they appear to gradually weaken farther along the jet length.  

In the southern arm, all the knots show both [Fe~II] and H$_2$ emission. HH81 and HH80, located towards the southern extremity of the jet are complex in nature which we attribute to thermal instabilities associated with low densities. Overall, the nature of shocks in the knots would suggest that densities of the ambient medium are relatively lower towards the southern arm as compared to the northern arm. This is validated by the fact that optical emission from HH81 and HH80 is visible due to lower extinction. This is also in accordance with conclusions of \citet{1993ApJ...416..208M} that the knots in the northern arm interact with the molecular cloud unlike the southern knots HH80 and HH81, which lie beyond the edge of the molecular cloud. A comparison of fields towards the northern and southern regions of the jet in the \textit{JHK} three-color composite image shown in Fig.~\ref{fig:three_color_comp} of Appendix~\ref{color_comp} indicates that the northern field displays a lower number of sources which are redder as compared to the southern field. 

In view of the fact that we detect both [Fe~II] and H$_2$ emissions in all the knots in the southern arm, we carefully analyze the variation of the corresponding $\mathscr{R}$ values with radial distance. The $\mathscr{R}$ values for all the knots are tabulated in Table~\ref{tab:flux}. One can discern that with an increase in radial distance from the central YSO, $\mathscr{R}$ decreases from Knot~5 to Knot~9 with a sharp decline in the value at Knot~7. The decrease in $\mathscr{R}$ with radial distance is likely to be due to a decrease in the strength of dissociative shocks. Such a decrease in $\mathscr{R}$ with radial distance has also been observed in other jet systems \citep{2000MNRAS.318..747D,2002A&A...393.1035N,2002ApJ...564..839L}. In addition, the lower densities towards Knots~8 and~9 could also result in a decrease in H$_2$ emission in these knots compared to the ones at smaller radial distances.

For each knot, we compare the NIR narrow-line observations with the radio emission from ionized gas, wherever available. We note that the knots giving rise to ionized gas emission are associated with [Fe~II] emission. However, the converse is not true, and the exceptions are Knot~5 (no 610~MHz emission) and Knot~7 where the [Fe~II] emission does not have a corresponding ionized gas emission. This is consistent with our assertion that the shock in this knot is J-type with a magnetic precursor. This is weaker than the strong J-type shock and it is highly likely that ionization of Fe is favorable against the ionization of hydrogen, as the former has lower ionization potential than the latter. This would explain the lack of ionized gas emission as well as the presence of [Fe~II] towards this knot.

The overall geometry of the jet on a large scale shows a curved S-shape centered on the MM1 core in IRAS 18162-2048. This S-shape of the distribution of knots was first noted in radio by \citet{1993ApJ...416..208M}. We have used NIR emission from multiple knots including those that emit in radio to confirm the curved morphology of the knot train. In addition, we note that a direct evidence of wiggling is seen in HH80 \citep{1993ApJ...416..208M}, where H$_2$ is linearly elongated along the lateral edge of the jet which provides support for a wiggling jet scenario. Similar geometries have been observed before in various other jets \citep{2002ApJ...564..839L, 1997AJ....114..265G, 1997AJ....114..280E}. Two possible scenarios can be considered (i) the jet interacting with dense ambient medium \citep{2002ApJ...564..839L}, or (ii) the wiggling of jet \citep{10.1093/mnras/260.1.163}. The symmetric shape is believed to result from the wiggling of jet due to precession \citep{2001A&A...372..899B, 2002ApJ...564..839L, Takami_2011}. The jet precession, in turn, can be attributed to 
(A) the presence of a companion, or to (B) asymmetry in the accretion of envelope into the disk.
The presence of a companion could lead to (a) the precession of the accretion disk due to the tidal interactions with a non-coplanar companion or to (b) the orbital motion of the jet source around a companion in orbit with the core. In case (B), the asymmetry could be due to a large difference in the orientation of magnetic and rotation axes of the core as demonstrated by simulations \citep{10.1093/mnras/stz740}. ALMA studies at a resolution of $40$~mas ($\sim56$~au) did not reveal the presence of a binary companion within MM1 core at these separations \citep{2019A&A...623L...8B}, and therefore higher resolution studies are required to draw valid conclusions regarding the presence of a binary companion.

\subsection{Knot luminosities and Mass-loss rates} \label{mass_loss}

The luminosities of knots in H$_2$ and [Fe~II] lines are calculated using dereddened fluxes and are in the range $10^{31}-10^{33}$~erg~s$^{-1}$, tabulated in Table~\ref{tab:lum_mass}. For this, we have assumed A$_v=30$~mag \citep{2001MNRAS.326..524D} for the knots close to the central region and in the northern arm (Knots~1, 2, 3, 4 and 5). \citet{rodriguez2019particle} had calculated A$_v=2.8$~mag and $3.4$~mag for HH80 and HH81 (Knots~8 and 9), respectively. For the knots at intermediate distances (Knots~6 and 7), we have assumed A$_v=15$~mag. We note that the ratio of [Fe~II] and H$_2$ luminosities $\bigg(\frac{L_{[Fe~II]}}{L_{H_2}}\bigg)$ are in the range $1.4 - 70.0$. This is higher than the values obtained for low mass class~0 and class~I YSOs for which $\frac{L_{[Fe~II]}}{L_{H_2}} \sim 10^{-2} - 10^{-1}$ \citep{2006A&A...449.1077C,2003A&A...397..693D}. We did not find this ratio for jets from massive YSOs in literature. We attempt to understand the observed $\frac{L_{[Fe~II]}}{L_{H_2}}$ ratio in terms of the dynamical age and evolutionary stage of the YSO driving the jet. The dynamical timescales of the HH80-81 radio jet is $\sim 10^{4}$~yr \citep{2012ApJ...758L..10M} while the associated outflows provide an estimate of $\sim 10^{6}$~yr \citep{10.1111/j.1365-2966.2004.07212.x}. As massive objects have shorter evolutionary timescales compared to their lower mass counterparts, this would suggest that the driving source is at a later evolutionary phase when compared to the low mass class 0/I sources of similar dynamical ages \citep{2000prpl.conf...59A,1992ApJ...384L..53B}. This indicates that the difference in $\frac{L_{[Fe~II]}}{L_{H_2}}$ is plausibly the effect of evolutionary phase of the driving source which manifests its influence on the jet physical properties. This is also corroborated by the large extent of the jet $\sim18$~pc. Another noteworthy point is that the ratio $\frac{L_{[Fe~II]}}{L_{H_2}} > 1$ implies a weakening of molecular line emission with the age of the driving source. This is expected because the clearing of the ambient medium by the previous mass-loss activities would result in lower densities and thus weak ambient magnetic field \citep{1997IAUS..182..181H} thereby favoring the dominance of dissociative J-type shocks over non-dissociative C-type shocks.

We also compute the mass-loss rates ($\dot{M}_{\text{jet}}$) of the knots using the mass ($M_{\text{jet}}$), tangential velocity ($v_\perp$) and sky-projected length ($l_\perp$), by using the following formula.

\begin{equation} \label{eq:emission}
\begin{split}
\dot{M}_{\text{jet}}& = M_{jet}\frac{v_\perp}{l_\perp}\\
 &  = \mu m_H L_{\text{[Fe~II]}} \bigg(h \nu A_i f_i \frac{Fe^{+}}{Fe} \bigg[\frac{Fe}{H}\bigg]\bigg)^{-1} \frac{v_\perp}{l_\perp}
\end{split}
\end{equation}

Here, $\mu = 1.24$, $m_H = 1.67\times 10^{-27}$~kg and $A_i = 0.00465$~s$^{-1}$ \citep{1988A&A...193..327N} are mean atomic weight, proton mass and radiative rate of the considered transition, respectively. We assume that the knot temperatures are $\sim 10^{4}$~K since radio emission is detected in the majority of our knots. For this temperature and electron densities of $\sim 10^{4}$~cm$^{-3}$ which is typical of protostellar jets \citep{2005A&A...441..159N,2013ApJ...778...71G,Takami_2006}, the fractional population of upper level of the transition $f_i = 0.01$ \citep{1994ApJ...436..292H} and $\frac{Fe^{+}}{Fe}\sim 0.25$ \citep{1994ApJS...93..485H}. We assume a gas phase abundance of iron $\bigg[\frac{Fe}{H}\bigg] = 3.09\times 10^{-5}$ which is the solar abundance \citep{2021A&A...653A.141A} assuming that there is no dust depletion. 

To get estimates of the velocities of knots, we utilize proper motion studies of HH80-81 from literature. Radio observations have shown that the tangential velocities of knots close to the YSO can be as high as 1000~km~s$^{-1}$ \citep{1993ApJ...416..208M,1995ApJ...449..184M}. Optical proper motion study by \citet{1998AJ....116.1940H} had revealed a range of velocities for the sub-knots of HH81 and HH80, and here we adopt an average velocity of 500~km~s$^{-1}$ for both these knots. We also assume $v_{\perp} = 500$~km~s$^{-1}$ for the knots at intermediate distances. The extent of [Fe~II] emission along the jet in each knot is adopted as the projected length ($l_{\perp}$).

The inferred mass-loss rates for the knots are in the range $3.0 \times 10^{-7} - 5.2 \times 10^{-5}$~M$_\odot$~yr$^{-1}$, tabulated in Table~\ref{tab:lum_mass}. For the knots close to the YSO, the values are consistent with those obtained from other tracers: $3\times10^{-5}$~M$_\odot$~yr$^{-1}$ from molecular gas using CO \citep{2019ApJ...871..141Q}, and $\sim10^{-5}$~M$_\odot$~yr$^{-1}$ using radio emission from ionised gas \citep{2012ApJ...752L..29C} towards the central region of the jet system. The mass-loss rates of HH80-81 jet are larger than those from jets of low-mass YSOs, $10^{-10}-10^{-7}$~M$_\odot$~yr$^{-1}$ \citep{2009ApJ...692....1D,2008A&A...487.1019G,10.1093/mnras/stw2296,2003A&A...397..693D}, but comparable to the [Fe~II] mass loss rate from few massive YSOs, $10^{-7}-10^{-4}$~M$_\odot$~yr$^{-1}$ \citep{2018A&A...616A.126F,2019NatCo..10.3630F,2008A&A...485..137C}. Assuming $\frac{\dot{M}_{\text{jet}}}{\dot{M}_{\text{acc}}}\sim 10\%$ \citep{2007IAUS..243..203C,2008A&A...479..503A}, we obtain $\dot{M}_{\text{acc}}$ of $\sim 10^{-6} - 10^{-4}$~M$_{\odot}$~yr$^{-1}$ for the knots. The accretion rates obtained from the knots close to the YSO are consistent with the observational and modeling estimates obtained towards the central region of this source \citep{2012ApJ...752L..29C,2020ApJ...888...41A} and is higher than the accretion rates observed towards low-intermediate mass protostars \citep{2013A&A...551A...5E,2009ApJS..181..321E,Calvet_2004,2014A&A...572A..62A,2020A&ARv..28....1L}. 

\subsection{The central region}

In this section, we examine the region close to the driving source and its immediate surroundings. The driving source of the jet, MM1 is not visible in the NIR J, H and K or the narrow-band images implying the highly embedded nature of the exciting YSO. A molecular outflow has been observed to be originating from MM1 with a position angle (PA) of 19$^{\circ}$ \citep{2019ApJ...871..141Q} and is found to be associated with the radio jet. Molecular line observations using CO (2-1), (3-2), (6-5) and (7-6) lines by them have revealed the red- and blue-shifted components of this outflow aligned along the direction of the radio jet. The outflow is marked as a white arrow pointing in the direction of the blue-shifted emission in Fig.~\ref{fig:color_comp}. In the vicinity of MM1, we also observe two arc-shaped structures in the west and south-west which resemble bow shocks; arcs 8 and 9 in the figure. These arcs appear to surround MM1 towards the western side and we believe that these could be bow-shocks associated with the walls of wind-swept wide angle cavities of strong winds  \citep{2002A&A...382.1021D} ejected by MM1.

Additional outflow components have been detected from the central region emanating towards north-east, north-west and south-east using sub-mm SiO(2-1), HCO$^+$(1-0), HCN(1-0) and CO(3-2) lines \citep{2013ApJ...778...72F}. These outflows are shown as magenta and green arrows in Fig.~\ref{fig:color_comp}. The lengths of the arrows indicate the approximate extent of the detected emission from the outflows. These outflows are believed to originate from MM2, which is $\sim 10^{4}$~au from MM1. MM2 comprises two cores, towards the east MM2(E) and west MM2(W) \citep{2019A&A...623L...8B}. The string of H$_2$ knots, in Reg~1 are aligned with the outflow in the south-east direction and we believe that these H$_2$ knots are shock-excited by the associated jet. The extent of the knots matches the extent of the detected outflow lobe. We note that these 5 knots trace a curve indicating precession of the outflow. 

There is a H$_2$ knot in Reg~6 which appears to lie on the opposite side of this outflow. Although no outflow lobe has been detected in this direction, we believe that this knot could plausibly be due to the other lobe of the same outflow that give rise to knots in Reg~1. Reg~4 has two H$_2$ knots that are aligned along the north-east SiO(2-1) outflow lobe. Again, we note a correspondence between the location of the knots and the extent of the outflow detected. Reg~2 has two H$_2$ knots which are located to the north of Reg~1 and south of Reg~4. It is difficult to comment on the likely outflow or exciting source responsible for this knot due to the cluster of sources present in the region. We also note arcs 11 and 12 facing inwards within the ellipse of Reg~7 appear to extend farther along the south-east of Reg~1. It is possible that they are bow-shocks associated with the same outflow. \citet{2001MNRAS.326..524D} had previously probed the central region (within 40$\arcsec$) using echelle spectroscopy and identified two bright H$_2$ knots and faint diffuse emission within the slit. The knots identified by them coincide with H$_2$ emission arcs that we discern towards the central region.

We also observe a number of H$_2$ arcs in Regs~3 and 5. These are seen to be surrounding the 8~$\mu$m nebulous regions which are located about $\sim20\arcsec$ from MM1. While it is possible that a few of these may be associated with outflows discussed above, we explore the possibility of YSOs in the vicinity as the source of excitation of these bow-shocks. Of the several YSOs identified in this region by \citet{2008ApJ...685.1005Q}, we find that two YSOs namely YSO1 and YSO2 are associated with MIR nebulosity as well as radio emission, see Fig.~\ref{H2_spitzer}. Fig.~\ref{H2_spitzer} shows that Reg~5 envelopes the H$_2$ arcs 1-7 and 8~$\mu$m nebulosity around YSO1. On the other hand, Reg~3 encloses H$_2$ arcs 10-13 and 8~$\mu$m nebulosity around YSO2. YSO1 and YSO2 have been identified as intermediate to massive YSOs by \citet{2008ApJ...685.1005Q} based on NIR and MIR CCDs. We have identified these objects as Stage~0/I objects by constructing their spectral energy distributions (SEDs) and fitting them with the radiative transfer models of \citet{2007ApJS..169..328R}. We thus believe that YSO1 and YSO2 could be driving strong winds that results in the formation of H$_2$ arcs in Reg 3 and 5, which surrounds these YSOs.

\section{Summary} \label{summary}

The impact of supersonic materials ejected from a rotating system on the ambient medium results in shock formation. For the first time, we present a comprehensive NIR/MIR view of shock formation associated with the HH80-81 jet in this paper. We discuss the knots generated by the HH80-81 jet in the context of strong, intermediate and weak shocks. We have utilized NIR H$_2$ and [Fe~II] emission lines to characterize the nature of the shocks at play in various knots. In the northern arm of the jet, we find that the knots closer to the central YSO show only [Fe~II] emission, the knots at intermediate distsnce show a combination of [Fe~II] and H$_2$ emission and the farther knots show only H$_2$ emission. From this, we infer that the shock strength is highest close to the driving source which gradually weakens with an increase in radial distance. In the southern arm, the shocks do not show such a transition but rather remain strong enough to excite [Fe~II] even in the farther knots of the jet, and thus all the knots in this direction show both H$_2$ and [Fe~II] emission. The knots in the southern arm have complex morphologies compared to the north and we attribute it to lower ambient density in the southern side. We determine the H$_2$ and [Fe~II] luminosities of all the knots to be in the range $0.5-13.2\times 10^{31}$ erg~s$^{-1}$ and $0.3-34.5\times 10^{32}$ erg~s$^{-1}$, respectively. For those knots in which [Fe~II] emission is detected, we obtain mass-loss rates of $\sim 3.0\times 10^{-7} - 5.2\times 10^{-5}$~M$_{\odot}$~yr$^{-1}$ which are consistent with the estimates from other massive protostars. We correlate the arcs and knots in H$_2$ emission towards the central region with multiple outflows detected as well as with YSOs in the vicinity.

\section*{Acknowledgment}
The NIR data presented in this paper are obtained using UKIRT and the instruments WFCAM and UIST. The WFCAM data were processed by Cambridge Astronomical Survey Unit (CASU). We thank the UKIRT team and CASU for carrying out the observations and providing us with the data products. This work is based [in part] on observations made with the Spitzer Space Telescope, which was operated by the Jet Propulsion Laboratory, California Institute of Technology under a contract with NASA. This research made use of data products from the Midcourse Space Experiment. Processing of the data was funded by the Ballistic Missile Defense Organization with additional support from NASA Office of Space Science. This research has also made use of the NASA/ IPAC Infrared Science Archive, which is operated by the Jet Propulsion Laboratory, California Institute of Technology, under contract with the National Aeronautics and Space Administration. We also thank the MHO catalogue service hosted by the University of Kent.

\bibliography{nir_jet}{}
\bibliographystyle{aasjournal}

\appendix

\section{Morphology of knots in narrow-line emission}

We examine and compare the morphologies of the nine knots in H$_2$ and [Fe~II] emission here.

\noindent \textit{Knot~1}: We have detected solely molecular H$_2$ emission towards this knot, [Fe~II] emission has not been detected. The morphology of the H$_2$ emission appears elongated and the emission appears to trace the edges of the jet with a lack of emission in the central region. The eastern edge appears to demonstrate a broken morphology. We do not observe any radio emission towards this knot.\\

\noindent \textit{Knot~2}: Similar to the case of Knot~1, we detect only molecular H$_2$ emission from this knot. The morphology of this knot is also elongated. However, unlike the case of Knot~1, the emission appears to emanate from the jet itself, although we do see hints of emission towards the edges. It is possible that we are viewing the front-edge of the cavity in H$_2$ towards us. Similar to the case of Knot~1, we do not observe any radio emission towards this knot.\\

\noindent \textit{Knot~3}: We detect both H$_2$ and [Fe~II] emission at distinct locations in this knot. The morphology of emission in both lines is very different. There are two [Fe~II] knots and two H$_2$ knots here. A comparison with radio emission towards this knot suggests that [Fe~II] knots trace the radio emission, while the H$_2$ knots lie to the south of the radio and [Fe~II] emission. \\

\noindent \textit{Knot~4}: This is the knot in the jet closest to the exciting source in the northern side. We detect only [Fe~II] emission towards this knot. The emission appears to have an elongated morphology along the jet axis with at least six [Fe~II] knot structures aligned almost linearly. The [Fe~II] emission traces the ionized gas distribution in this case while the H$_2$ features appear to trace the cooler edges. \\

\noindent \textit{Knot~5}: This is the knot in the jet that is closest to the exciting source in the southern side. Here we observe two [Fe~II] and three H$_2$ knots, lying at distinct locations. The H$_2$ knots are located in the outer direction of the propagating jet. We observe that the knots appear to be aligned along the jet direction.
Here, ionized gas emission is detected at 20~cm \citep{1993ApJ...416..208M}, overlapping with [Fe~II] and H$_2$ emission. The radio emission appears to be more extended than the NIR emission. However, radio emission is not detected at low radio frequencies of 610 and 325~MHz. This suggests that the 20~cm radio emission can plausibly be attributed to thermal excitation rather than non-thermal emission.\\

\noindent \textit{Knot~6}: We observe both [Fe~II] and H$_2$ emission towards this knot, with an overlapping region of emission. The [Fe~II] emission morphology resembles an arc with the head pointing in the outward jet direction while the H$_2$ emission is located towards the western wing of the arc with a region of overlap. Here the ionized gas emission overlaps with [Fe~II] and H$_2$ emission. \\

\noindent \textit{Knot~7}: We observe both H$_2$ and [Fe~II] emission towards this knot, which displays a complex morphology. The H$_2$ emission appears to resemble a broken bow morphology. Towards the north, we notice two arc-shaped structures with no emission in the center, which could be part of a broken bubble. Elongated emission is observed towards the south-east of the bubble-like structure. Two faint knots of [Fe~II] emission are located towards the interior of the H$_2$ emission. No radio emission is detected towards this knot. \\

\noindent \textit{Knot~8}: This knot corresponds to HH81 and we observe [Fe~II] as well as H$_2$ emission towards this knot. For comparison, we have included optical emission from HST in the filter H$\alpha$+[NII]. We note that there is an overlap between the optical and [Fe~II] emission, visible as cyan in Fig.~\ref{fig:color_comp_full} where optical emission is represented in blue. This emission reveals the presence of various sub-knots within the main knot. We observe an elongated emission and a compact knot in [Fe~II] towards the south of the optical emission. \citet{1998AJ....116.1940H} have identified two sub-knots structures namely A and B in the optical emission and both of these overlap with two of the [Fe~II] knots in the north. The other two [Fe~II] knots and the two H$_2$ knots do not have any corresponding optical emission. Towards the north and east of the optical emission, we observe two arcs in the H$_2$ emission. There is a marginal overlap between the H$_2$ and [Fe~II]/optical emission. The orientation of the [Fe~II]/optical bow as well as the northern H$_2$ bow are in a direction that is opposite to the jet propagation. There is no overlap between the [Fe~II] and H$_2$ emission. The associated ionized gas emission has spatial overlap with [Fe~II] and optical emission, but no overlap with H$_2$ emission. A comparison with soft X-ray emission towards this knot as discussed by \citet{2004ApJ...605..259P} in their Fig.~3 suggests that the X-ray emission lies in the vicinity of Knot~A to the south-west.\\

\noindent \textit{Knot~9}: This knot corresponds to HH80 and we observe [Fe~II] as well as H$_2$ emission towards this knot. Similar to the case of HH81 we include the optical emission from HST for a comparative study. \citet{1998AJ....116.1940H} have identified 12 sub-knots in the optical emission and labeled them in the order from A to L, shown in Fig.~\ref{fig:color_comp_full}. For sub-knots A to E, G and H, the optical emission and [Fe~II] emission overlap, which is visible as cyan in the figure. Knot~F is located above the corresponding optical sub-knot. For sub-knots I, J, K and L, we do not detect associated [Fe~II] emission. The H$_2$ emission, on the other hand, has a distinct location compared to both [Fe~II]/optical emission without any overlap. We observe four H$_2$ sub-knots, of which three are linearly arranged along the direction of jet propagation. However, this arrangement is outward of the [Fe~II]/optical emission towards the east and appears to be tracing the walls of the jet cavity. One H$_2$ sub-knot is located towards the north-eastern edge of the sub-knot~C. However, this extends outwards farther than the [Fe~II]/optical emission. Similar to HH81, there is no overlap of ionized gas emission in radio with H$_2$ emission. There appears to be an overlap of [Fe~II] emission with radio towards sub-knots~A to E, however, we cannot comment on the exact correspondence with sub-knots due to the relatively large beam-size of radio observations. For sub-knots~F, G and H which display [Fe~II] emission, there is no associated ionized gas emission. Soft X-ray emission has also been detected towards sub-knots~A, C and G by \citet{rodriguez2019particle}. This suggests the presence of hard shocks traveling at speeds greater than 300~km~s$^{-1}$ \citep{Bally_2002}. \\
 
Amongst the knots observed in NIR, all knots other than Knots~7, 8 and 9 show an elongated morphology along the jet axis. This is explicable on the basis of excitation in the immediate vicinity of the jet. Knots~7, 8 and 9, on the other hand, display a complex fragmented morphology.

\section{Evidence for denser cloud medium the northern arm of the jet} \label{color_comp}

\begin{figure*}
\hspace*{-1.5cm}
\centering
	\includegraphics[trim={0cm 0cm 0cm 0cm},clip, scale = 0.8]{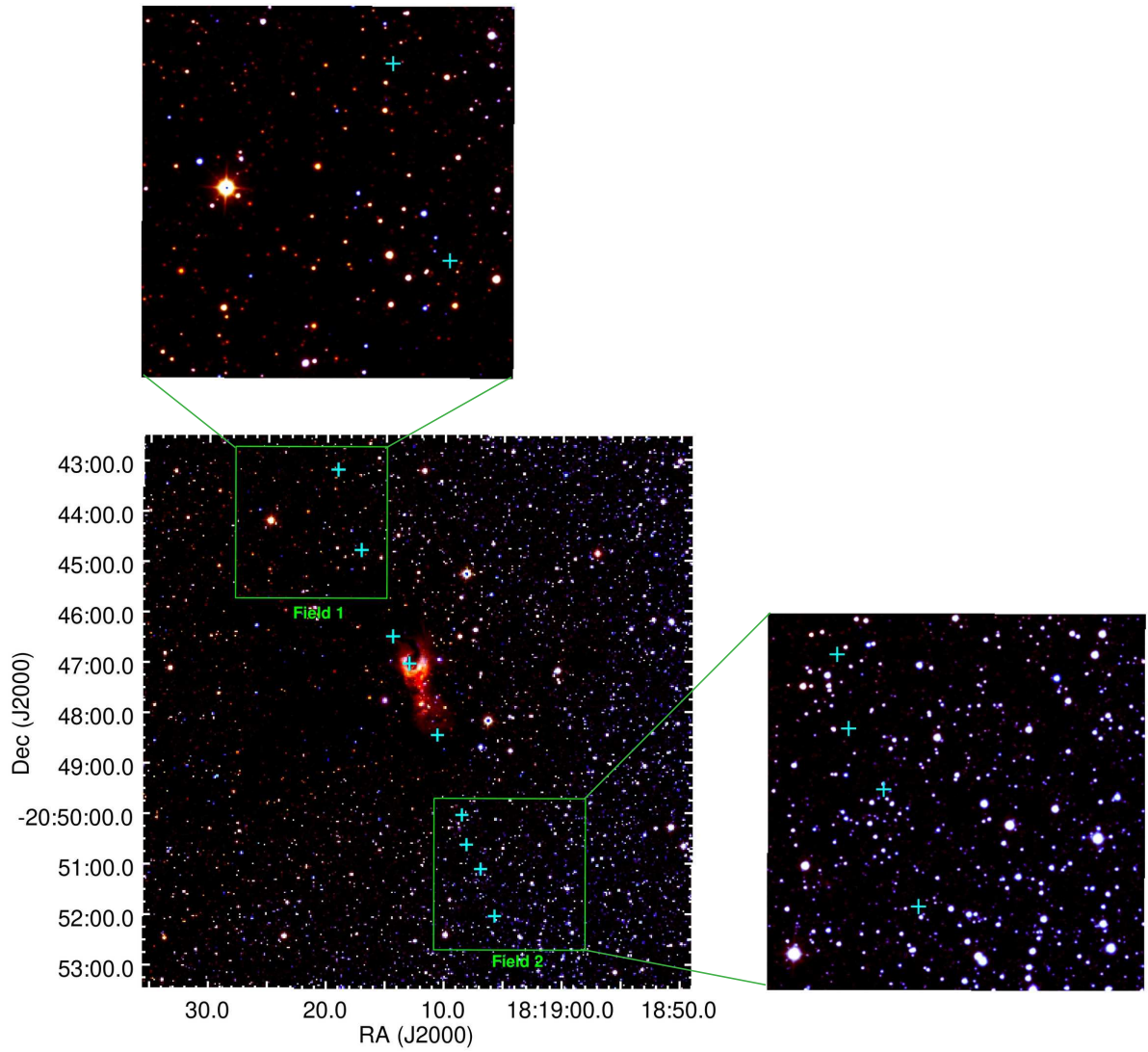} %
    \caption{\textit{JHK} color-composite image of the HH80-81 jet, with K band emission shown in red, H band emission in green and J band in blue. Knots~1 $-$ 9 are marked with cyan crosses. The green boxes marked as Fields~1 and 2 show two regions towards the northern and southern arms of the jet, respectively. }
    \label{fig:three_color_comp}
\end{figure*}

Fig.~\ref{fig:three_color_comp} shows the three-color \textit{JHK} composite image of the HH80-81 jet region. In the figure, the K band emission is shown in red, the H band emission in green and the J band in blue. Knots~1 $-$ 9 are marked with cyan crosses. The regions enclosed in boxes with green outline and marked as Field~1 and 2 represents two fields of sizes 3$\arcmin$ $\times$ 3$\arcmin$. These boxes have been used to compare the overall source density and reddening towards the northern arm and southern arm of the jet, respectively. We note that there are fewer sources in Field~1 as compared to Field~2. In addition, the sources in Field~1 are more reddened sources in comparison to those in Field~2. This implies the presence of a dense molecular cloud towards the northern arm. This is also consistent with the conclusions of \citet{1993ApJ...416..208M} that the knots in the northern arm interact with the molecular cloud unlike the southern knots HH80 and HH81, which are detected in optical as they lie beyond the edge of the molecular cloud.

\end{document}